\documentclass{article}
\usepackage{fullpage,url,amssymb,hyperref}
\usepackage{graphicx}
\usepackage[style=apa, backend=biber]{biblatex}
\begin{document}
\title{Impact of introducing ``Informatics I'' to the common university entrance examination in Japan: a longitudinal study on students' perceptions of their information-related knowledge and skills from 2006 to 2026\footnote{This paper contains parts of the following two articles, both in Japanese: A. Morihata ``hongaku 1 nensei no jouhoukei ginou / chisiki no suii ni tsuite'', Bulletin of Education and Research Data Analysis Office, College of Arts and Sciences the University of Tokyo, Vol. 6, pp. 3--16, 2025; and  A. Morihata ``Zoku: hongaku 1 nensei no jouhoukei ginou / chisiki no suii ni tsuite --- nyushi no eikyo wo chushin ni---'', Bulletin of Education and Research Data Analysis Office, College of Arts and Sciences the University of Tokyo, Vol. 7, pp. 1--15, 2026. This paper substantially extends them in the following aspects: the addition of FY 2026 data, a literature review, and a substantial deepening of the discussion. In particular, the preceding articles neither conducted statistical tests nor applied the theoretical framework of the washback effect.}}
\author{Akimasa Morihata (University of Tokyo)}
\date{}
\maketitle{}

\begin{abstract}
Despite the recent intensive development of secondary education
curricula and assessments in informatics, the impact of assessments
has not been well studied in this field. Since informatics education
covers a diverse range of content, from computer science
knowledge to ICT skills, careful consideration is needed to prevent
assessments from distorting education.
This study investigates the
impact of introducing ``Informatics I'' into the Common Test for
University Admissions in Japan, as an example of a large-scale,
standardized, high-stakes assessment in 2025. As the data source for
this analysis, this study uses a questionnaire that has been administered
every year from 2006 to 2026 to all first-year students at 
the University of Tokyo, a major comprehensive national university in Japan. 
The questionnaire asks students
for their self-perceptions of the information-related knowledge and skills they studied and acquired in high school. 
Using these data, we conduct a longitudinal study of the 2013 curriculum
reform, the 2022 reform, and the introduction of the new entrance
examination. We attempt to isolate the impact of the entrance
examination through two comparisons: between the 2013 curriculum
reform (which involved no change to entrance examinations) and the
2022 reform (which led to the new entrance examination in 2025); and
between direct-entry and gap-year students among those entering in
2025, who followed different curricula but took the new
examination. We use the theoretical framework of the washback effect
as a lens for interpreting these differences. We found that (1) the
2013 curriculum reform produced no discontinuity in students'
perceptions, whereas (2) the introduction of the new entrance
examination in 2025 produced a sharp change, particularly in the
proportion of students reporting acquisition of computer-science topics; and (3) this change is too large to be
interpreted as a gain in proficiency, and is better understood as a
shift in students' criteria for judging acquisition. 
\end{abstract}

\section{Introduction}
Over the past quarter century, secondary education curricula in the informatics field have been developed rapidly.
Representative curriculum standards include
the K-12 Computer Science Standards\footnote{\url{https://csteachers.org/k12standards/}} by the Computer Science Teachers Association and
the ISTE standards\footnote{\url{https://iste.org/standards/}} by the International Society for Technology in Education.
Informatics curricula have been introduced into primary and secondary education in many countries \parencite{Thompson13,Bell14,Brown14,Hubwieser2015,Syslo15,Hong16,Sentance17,Moller18,Larke19,Sentance22,Faherty23}.

Compared with the rich discussion of curriculum design in informatics education,
relatively few studies have addressed student assessment \parencite{Mattern11,Bienkowski15,Scanlon21,Munoz23}; 
especially, those evaluating \emph{whether assessment improves teaching and learning} remain scarce.
In education, leaving certificates and entrance examinations often determine what is taught and what is learned.
Informatics education includes diverse subjects, such as computational thinking, information design, ICT skills, computer science, and computer system architecture, and accordingly calls for a tailored approach to assessment depending on the subject. Careful consideration is therefore needed to prevent assessment from distorting education.

This paper reports the case of Japan, where ``Informatics I'' was newly introduced as a subject of the nationwide standardized university entrance examination in 2025.
This examination is exceptional in the informatics field in that about half of all university applicants take it regardless of their intended major (that is, not only those in STEAM fields).
We analyze the impact of this new examination on high-school teaching and learning, based on a questionnaire administered to first-year students at the University of Tokyo, a major comprehensive national university in Japan. The questionnaire, which has been administered every year from 2006 to the present, asks students to report their self-perceptions of the information-related knowledge and skills they studied and acquired during high school. 

Through a longitudinal study using these 21 years of data, we attempt to identify the impact of the new entrance examination through the two key comparisons.
The first is the comparison between the curriculum reform in 2013 (without reform of entrance examinations) and the curriculum reform in 2022, which led to the introduction of the new entrance examination in 2025.
The second is the comparison among three groups: students entering in 2024 (who studied the curriculum before 2022 and took no entrance examination in informatics),
gap-year students among those entering in 2025 (who studied the curriculum before 2022 and took the new examination), and direct-entry students (who studied the curriculum from 2022 and took the new examination).
These comparisons aim to isolate the impact of the new examination from that of the curriculum reform. 
Furthermore, to interpret the impact accurately, we adopt as a lens
the theoretical framework of the washback effect \parencite{Alderson93,Cheng04,Green13}, in particular findings on \emph{learners' perception} \parencite{Xie25}.

The results of this study will provide valuable insights, not only for Japan but also worldwide, into the further development of informatics curricula, notably the design of large-scale, systematic assessment in light of the ``Computer Science for All'' trend.

\section{Background}
In Japan, compulsory Informatics in high school began in 2003, reflecting the spread of PCs and the Internet. There were three subjects, ``Informatics A'', ``Informatics B'', and ``Informatics C'',  loosely corresponding to ICT literacy, the K-12 Computer Science Framework, and the ISTE standards, respectively. We call this period the first period. 

When the first-period curriculum was designed, computers and the Internet were advanced technologies. They quickly became, however, technologies that everyone had to handle well. At the same time, ICT literacy education was steadily introduced into primary and secondary education. In light of this situation, from 2013, Informatics was reorganized into two subjects, ``Information Science'' and ``Information and Society'', loosely corresponding to Informatics B and Informatics C, respectively. We call this period the second period.

The intention of offering multiple subjects was to provide informatics education suited to the circumstances of each high school. In reality, however, about 80\% of high schools chose Informatics A as their required subject in the first period \parencite{Ikuta08}, and nearly 80\% of high school students studied Information and Society in the second period \parencite{mext15}. 
This was recognized as problematic, because students would not acquire the computer science proficiency needed to meet the large demand for ICT professionals. As a result, from 2022 the compulsory Informatics subjects were unified into ``Informatics I''. The content of ``Informatics I'' includes the two second-period subjects, and content related to data science. This history is also discussed in detail by \textcite{Hagiya15}, \textcite{Kanemune17}, and \textcite{Nakayama18}.

In parallel with this, the introduction of Informatics to university entrance examinations was discussed \parencite{Hagiya22,Yoshida23}. University entrance examinations in Japan are so high-stakes that they are even regarded as the most important examination in one's life. It was therefore argued that, unless Informatics was included in university entrance examinations, high school teachers and students would not teach or study it seriously.

From 2025, when the students who had studied under the third-period curriculum took university entrance examinations\footnote{High school in Japan lasts three years, beginning in April and ending in March. Students who entered high school in April 2022 took the CT in January 2025 and graduated in March 2025.}, ``Informatics I'' was added to the Common Test for University Admissions (hereafter, CT) of the National Center for University Entrance Examinations. The CT is effectively mandatory for university applicants in Japan who apply through the general admission track: it serves as a screening gate, especially at national and public universities, which include most of the top-ranked universities in Japan. Many universities and faculties required the newly introduced CT ``Informatics I''. In 2026, among approximately 620,000 students entering university, 460,000 students took the CT, and more than 300,000 students took CT ``Informatics I''.

Note that the situation regarding CT ``Informatics I'' is different from that of school-leaving certificates and university entrance examinations in other countries, such as the AP CSP test in the USA, NSI in France, A-level in the UK, the leaving certificate in Ireland, the Gaokao in China, the National High School Exam in Vietnam, and NZQA in New Zealand. In these countries, those who take the informatics test are in principle limited to students going on to STEAM fields or those who actively choose to take it.

\section{Literature Review}
Many studies have reported on the designs of informatics curricula and their reform, including New Zealand \parencite{Thompson13, Bell14}, the UK \parencite{Brown14,Sentance17,Larke19}, Poland \parencite{Syslo15}, USA \parencite{Hong16}, Wales \parencite{Moller18}, and Ireland \parencite{Sentance22,Faherty23}. Regarding assessment, they have noted the importance of assessment for implementing the curriculum \parencite{Larke19}, the challenge of designing appropriate assessment \parencite{Munoz23}, and teachers' demand for guidance on assessment \parencite{Sentance17}.

Only a few studies examined the impact of assessment in informatics education in depth, and \textcite{Mattern11} and \textcite{Scanlon21} are notable exceptions. \textcite{Mattern11} revealed a strong link between taking an AP Exam in computer science and majoring in computer science in college. \textcite{Scanlon21} conducted interviews with teachers and students and reported several findings: teachers were concerned that assessment would make students focus on particular activities and lead to biased learning; students' perceptions differed from teachers' expectations (for example, students regarded practical activities as more effective for learning theoretical knowledge than teachers did); and assessment had a positive effect by motivating students.

In light of the literature reviewed above, studies on the impact of assessment in informatics education remain insufficient.
In particular, we found no macro-level analysis of the impact of large-scale high-stakes paper tests.
Realizing ``Computer Science for All'' naturally requires standardized large-scale assessment, for which a large-scale paper test is an obvious candidate. However, such a test may give rise to an excessive focus on certain domains and approaches to learning, as \textcite{Scanlon21} pointed out.

The impact of tests on education and learning is called the washback effect\footnote{It is sometimes called the ``backwash effect''.} \parencite{Alderson93,Cheng04, Green13}. Hughes formulated the washback effect as a relationship among participants, processes, and products \parencite{Hughes93,Bailey96}. \textcite{Alderson93} refined this into 15 hypotheses and posed the questions of which aspects of teachers and learners are affected by washback and how, in what direction (i.e., positive or negative), to what extent, and with what outcome. In response to these questions, numerous studies have demonstrated various washback effects depending on the context \parencite{Cheng04,Green13}.

Studies that examine the washback effect on the perceptions of learners (test-takers) are relatively few \parencite{Cheng98,Qi07,Cheng11,Li12,Xie13,Pan14,Xie15,Zhang21,Dong22,Tsang22}.
Moreover, compared with the impact on teachers, the impact on learners is known to be complex and context-dependent. \textcite{Cheng98, Qi07} observed that what students learned did not change substantially---at least not in the way the test developers had intended. \textcite{Li12} reported that the test improved students' self-efficacy. These washback effects are influenced by learners' proficiency in the subject concerned \parencite{Cheng11,Pan14}, perceptions of test importance \parencite{Xie13} and of test design \parencite{Xie15}, and even by home region and high school \parencite{Zhang21,Dong22} and social networks \parencite{Tsang22}.

We cannot directly infer, from previous studies, the washback effect caused by the introduction of CT ``Informatics I''. First, the introduction of a new high-stakes test can produce an impact different from that of a change in content or in institutional arrangements. Introduction of a new large high-stakes test is rare; notable examples include those on the introduction of university exit English tests in Taiwan \parencite{Pan14, Hsieh17}. Second, CT ``Informatics I'' was introduced when almost no comparable test existed, and hence it is doubtful whether the case is the same as that of the new English tests. Third, studies of the washback effect in contexts other than foreign language learning are limited \parencite{Rodriguez16, Qureshi18,Fernández22}. To the author's knowledge, the only study covering the informatics field is a relatively small-scale one by \textcite{Jensen26}.

\section{Current Study}
The primary research question (RQ) of this study concerns the impact of the introduction of CT ``Informatics I''.
However, it appears to be difficult to isolate the impact of the CT from that of the curriculum reform.
In principle, students who took the CT studied under the third-period curriculum, whereas those who did not studied under the second-period curriculum.

To address this problem, as discussed in Introduction,
this study attempts to isolate the impacts of the CT according to the following strategy.
First, we focus on 2016, the year when second-period students began to enter,
and estimate the magnitude of the impact of the curriculum reform (without the introduction of the CT)
by comparing it with periods in which no curriculum reform was carried out.
Second, by contrasting this with 2025, the year when third-period students began to enter,
we roughly analyze the impact of the introduction of the CT.
Third, we classify the students entering in 2025 into direct-entry students and gap-year students (hereafter, DE and GY, respectively), both of whom took the CT but studied different curricula.
We analyze the impact of the introduction of the CT in more detail
by contrasting them with the students entering in the preceding and following years, in particular with the students entering in 2024, who followed the same curriculum as GY but did not take the CT.
To improve accuracy, we interpret the analysis results through the lens of washback effect studies.

We summarize the operational RQs as follows.
\begin{description}
\item[RQ1]  How have Japanese students' perceptions of the informatics field changed during the first period and the second period?
\item[RQ2]  Did the curriculum reform in 2022 and the introduction of the CT in 2025 produce a discontinuous change in students' perceptions of the informatics field?
\item[RQ3] To what extent do perceptions of the informatics field differ between students who followed the same curriculum but were subject to different entrance examination systems, and between those entering in the same year but following different curricula?
\end{description}

We approach these RQs using the ``Survey on the Study of High School Informatics\footnote{The questionnaire items and  summary statistic are available from \url{https://sites.google.com/site/iebtokyouniv/home/edu/information/}.}'' at the University of Tokyo.
This survey has been administered to first-year students every year since 2006, when students who had studied high school Informatics began to enter.
Among the questionnaire items of the survey, we mainly deal with those that ask about the knowledge and skills studied and acquired. Hereafter, we call the proportions of students who reported that they had ``studied'' or ``acquired'' a topic the ``study rate'' and the ``acquisition rate'', respectively.

Note that the answers report students' perceptions. For example, when students say that they studied \textsf{SpreadSheet}, some may have in mind creating graphs and scatter plots from data using spreadsheet software, while others may have in mind automating data analysis using functions and macros. Moreover, it depends on the respondent how much learning and understanding of a topic the student's answer of ``studied'' or ``acquired'' indicates. Furthermore, the answer ``studied'' does not even indicate the fact that the student was taught the topic in a high school class. The student may have forgotten the class content, or conversely may believe that they were taught the topic when in fact they were not. Also, the answer ``acquired'' does not mean that the student would get a high score on an objective test.

In this study, we turn this point to our advantage for analyzing the impact of the CT. We isolate the impact of the entrance examination by comparing students' perceptions with the actual curriculum. In particular, if students' perceptions changed even though the curriculum did not change, that strongly suggests a washback effect.

The same survey has been analyzed by \parencite{Yama26}. He reported the twenty-year (except for 2026) trends in the Informatics subject students studied in high school, as well as the studied and acquired rates. He conjectured that the abrupt change in 2025 reflects a change in their self-evaluation criteria rather than in their proficiency. This study shares the same observation but goes further. As discussed above, we isolate the effect of the CT introduction and interpret the results through the lens of the washback effect.

\subsection{Details of the Survey}\label{subsect:survey}
First-year students at the University of Tokyo consist of about 1250 students in humanities and social sciences (hereafter, HS) and about 1850 students in STEAM, about 3100 in total.
DE account for $70\sim 80\%$ and GY for $20\sim 30\%$. Hence, even when a curriculum reform occurs, its effect appears with a delay for the portion corresponding to the gap-year students. Most gap-year students spend one year, some spend two years, and very few spend three or more.

The survey analyzed here is administered on the first day of ``Informatics'', a compulsory subject for all first-year students. All enrolled students are asked to respond. This survey is fully anonymous and contains no information directly identifying individuals. Responding is voluntary, and students are informed that it does not affect their grades. Use of the survey results for this study has been approved by the Research Ethics Committee for Experimental Research on Human Subjects, the University of Tokyo (approval number: 1141).

We analyze the 13 question items in Table \ref{table:items1}. We refer to each item by the wording in the ``abbreviation'' column.
Table \ref{table:items2} shows whether the Courses of Study for each subject contain a description related to each item. For reference, it also shows the correspondence with the Core Concepts of the K–12 Computer Science Framework\footnote{\url{https://k12cs.org/}}. One can see that the curriculum has shifted from Informatics A, which centered on ICT literacy, to a curriculum centered on the computer science domain similar to those in K-12.

According to whether they are described in the Courses of Study for Informatics I, we classify these 13 items into ``Literacy'', consisting of the eight items \textsf{WordProcessing}, \textsf{SpreadSheet}, \textsf{Presentation}, \textsf{Email}, \textsf{WebSearch}, \textsf{Typing}, \textsf{MultiMedia}, and \textsf{WebPage}, and ``CS'', consisting of the five items \textsf{Programming}, \textsf{ComputerSystem}, \textsf{Simulation}, \textsf{Database}, and \textsf{Ethics}. This classification reflects the fact that the subject of this paper is the ``Informatics I'' entrance examination. ``CS'' corresponds to the items likely to be examined, and ``Literacy'' to those unlikely to be examined.

The question items were the same up to 2024. The wording was modified slightly from 2025 in order to align the items with a nationwide survey by the Information Processing Society of Japan \parencite{OSNN25}. The most significant change is the separation of \textsf{ComputerSystem} into ``Computers and digital representation'' and ``Information and communication networks''. For this item, we use the average value of the two items for 2025 and 2026. We note that the two items showed very similar trends. The possible effects of the other modifications are discussed in Section \ref{subsection:limitation}.

\begin{table}[!t]
\centering\small
\caption{Skills and knowledge covered in the Survey on the Study of High School Informatics}\label{table:items1}\smallskip
\begin{tabular}{l|l}
\hline
Item & Abbreviation \\
\hline
Basic operations of word processing software & \textsf{WordProcessing} \\
Basic operations of spreadsheet software & \textsf{SpreadSheet} \\
Basic operations of presentation software & \textsf{Presentation} \\
Basic operations and appropriate behaviors concerning email & \textsf{Email} \\
Web search & \textsf{WebSearch} \\
Touch typing (typing without looking at the keyboard) & \textsf{Typing} \\
Programming & \textsf{Programming}\\
How computers and networks work & \textsf{ComputerSystem} \\
Modeling and simulation & \textsf{Simulation}\\
Databases & \textsf{Database} \\
Image processing and multimedia & \textsf{MultiMedia} \\
Web-page creation & \textsf{WebPage} \\
Copyright, personal information protection, and privacy & \textsf{Ethics} \\
\hline
\end{tabular}
\bigskip\bigskip

\caption{Descriptions of the 13 items in the Courses of Study and their correspondence with the Core Concepts of  K-12. Inf. in the column headings abbreviates Informatics or Information.}\label{table:items2}\smallskip
\begin{tabular}{c|ccc|cc|c|c}
\hline
Abbreviation & Inf. A & Inf. B & Inf. C & Inf. Science & Inf. and Society & Inf. I & K-12\\
\hline
\textsf{WordProcessing} & \checkmark & & & & & \\
\textsf{SpreadSheet} & \checkmark &  & \checkmark & & & \\
\textsf{Presentation} &\checkmark  &  &  & &  & \\
\textsf{Email} &  &\checkmark  & & & \checkmark & \\
\textsf{WebSearch} & \checkmark &  &  & & & \\
\textsf{Typing} & & & & & & \\
\textsf{Programming} & & \checkmark & & \checkmark & & \checkmark & C4\\
\textsf{ComputerSystem} & & \checkmark & & \checkmark & \checkmark & \checkmark & C1, C2\\
\textsf{Simulation} & & \checkmark & & \checkmark & & \checkmark & C2\\
\textsf{Database} & \checkmark & \checkmark &  & \checkmark & & \checkmark & C3\\
\textsf{MultiMedia} & \checkmark &\checkmark & \checkmark & & & \\
\textsf{WebPage} & & & & & \checkmark & \\
\textsf{Ethics} & \checkmark &  & \checkmark & & \checkmark & \checkmark & C5\\
\hline
\end{tabular}
\end{table}

\subsection{Possibility of Bias}\label{subsect:bias}
The University of Tokyo adopted CT ``Informatics I'' in all admission categories from the 2025 entrance examination. This introduces a natural selection bias: students with good Informatics I scores are more likely to be admitted. However, this bias is unlikely to be large. Informatics I accounts for only 2\% of the total score of the entrance examination. Moreover, students entering the University of Tokyo commonly obtain high CT scores (typically 80\% or higher), and hence the spread of their Informatics I scores is confined to a narrow range. The actual score difference produced by Informatics I should therefore be about 0.5\% of the total score of the entrance examination.

\begin{figure}[tb]
\includegraphics[width=\linewidth]{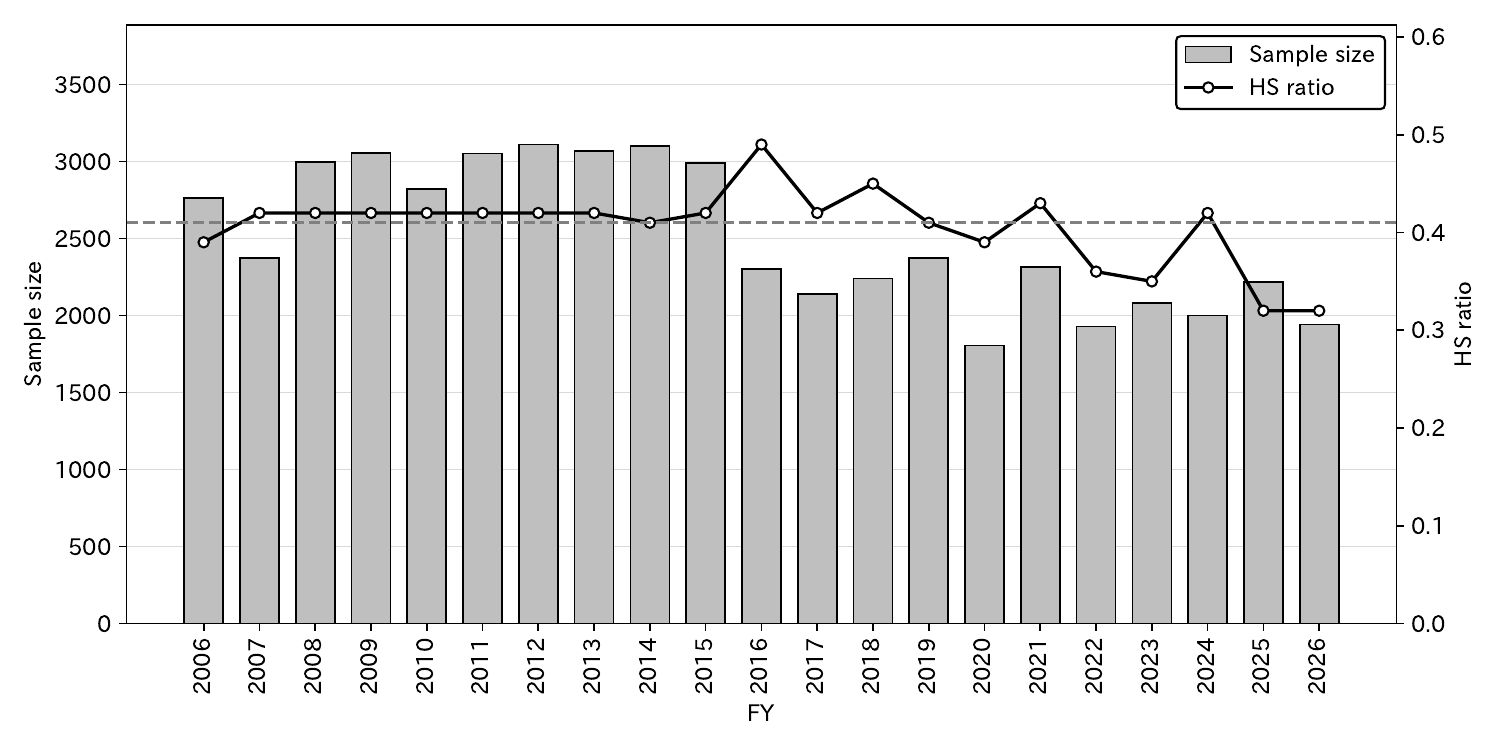}
\caption{The number of respondents and the proportion of HS students by year. The dotted line indicates the proportion of HS students among all entering students.}\label{fig:num_ans}
\end{figure}

Next, we consider non-response bias. Figure \ref{fig:num_ans} shows the change in the number of respondents. It also shows, as a proxy variable for bias, the proportion of respondents whose Admission Examination Category is HS. We selected this proxy variable because perceptions of the informatics field may correlate with Admission Examination Categories.

The number of respondents was high up to 2015 because paper questionnaires were distributed. It declined from 2016 because responses were collected online through the learning management system. Even in the year that had the fewest respondents, more than half of the entering students responded.

We next examine the biases observed via the proportion of HS students. In most years the bias is slight, but a small bias is seen in 2016, 2025, and 2026 (Cram\'{e}r's $V$ is $0.18$, $0.16$, and $0.18$ for 2016, 2025, and 2026, respectively). The possible effects of these biases are discussed later in Section \ref{subsection:limitation}.

\section{Results}
\subsection{Trends in Study Rates and Acquisition Rates}
\begin{figure}[p]
\centering
\includegraphics[width=.8\linewidth]{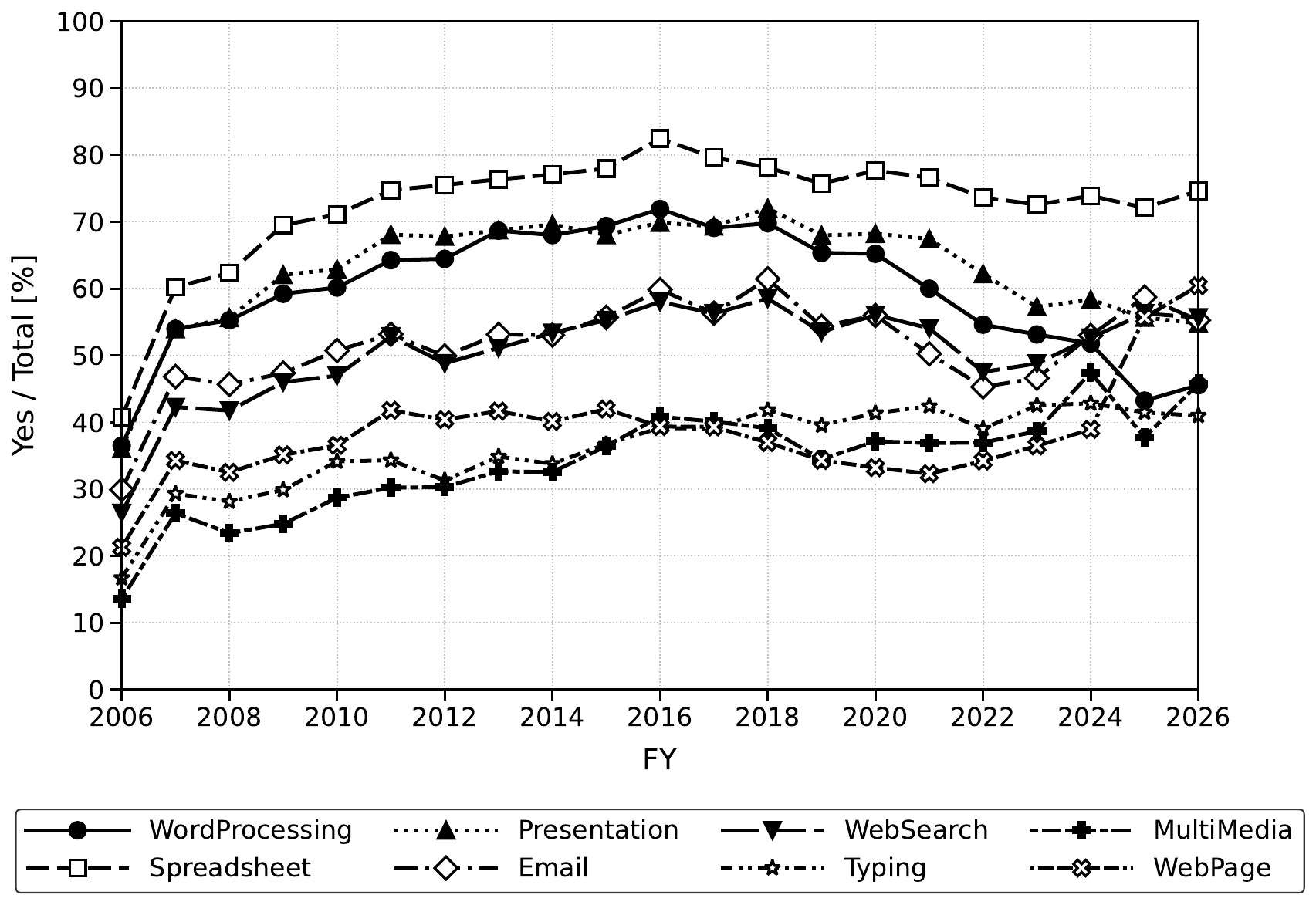}
\bigskip\bigskip\\

\includegraphics[width=.8\linewidth]{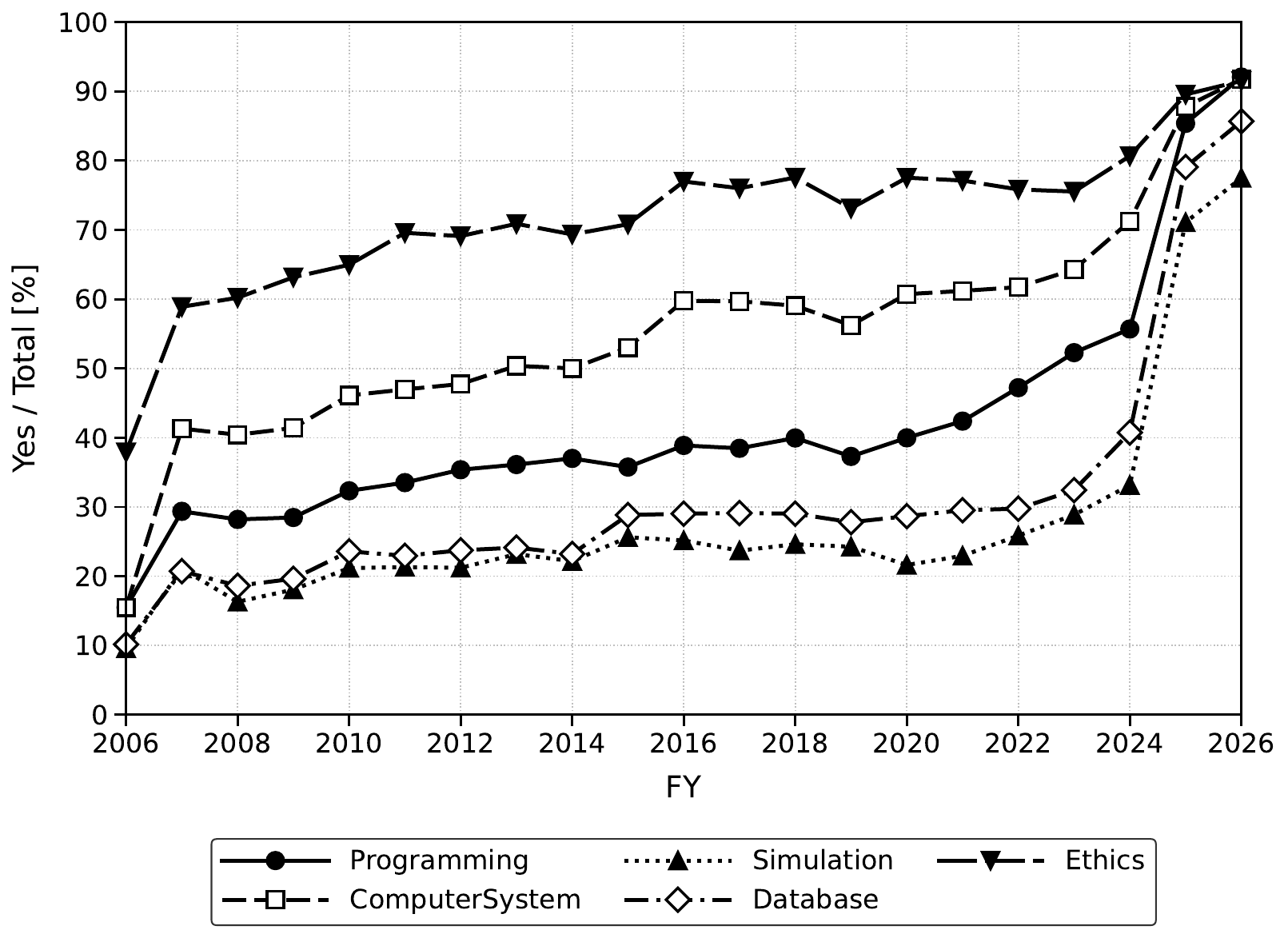}
\caption{Trends in study rates: (upper) literacy items, (lower) CS items. The $95\%$ confidence intervals based on the finite population correction are at most about $\pm 2\%$ in absolute value, a width smaller than the markers, and are therefore omitted.}\label{fig:learn}\bigskip
\end{figure}
\begin{figure}[p]
\centering
\includegraphics[width=.8\linewidth]{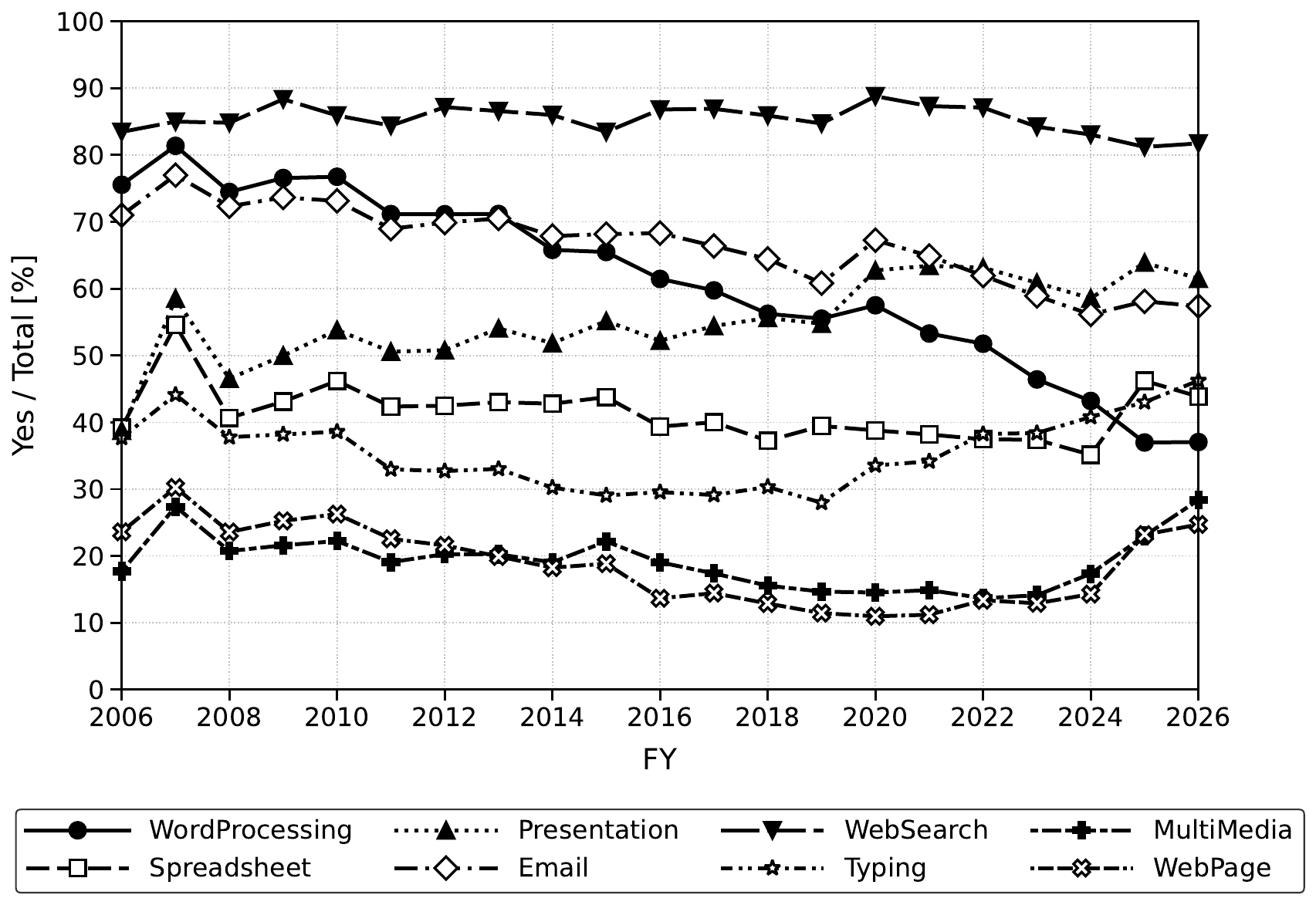}
\bigskip\bigskip\\

\includegraphics[width=.8\linewidth]{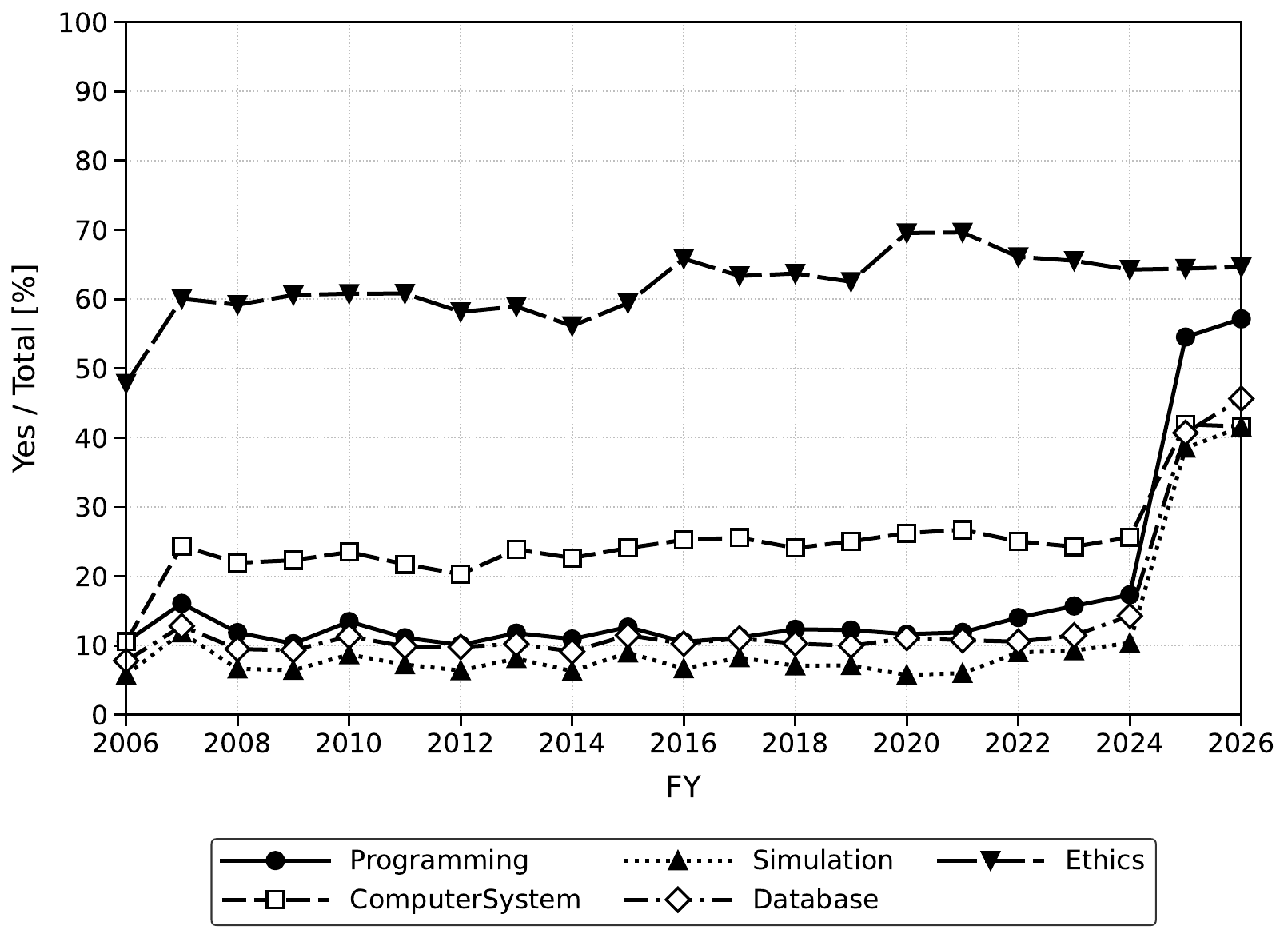}
\caption{Trends in acquisition rates: (upper) literacy items, (lower) CS items. The $95\%$ confidence intervals based on the finite population correction are at most about $\pm 2\%$ in absolute value, a width smaller than the markers, and are therefore omitted.}\label{fig:studied}
\end{figure}

Figures~\ref{fig:learn} and \ref{fig:studied} show the trends in study rates and acquisition rates, respectively.

The study rates for the literacy items tended to increase gradually in the first period and to decline slowly from the second period, notably for \textsf{WordProcessing} and \textsf{Presentation}. Even so, they maintained high values overall. In the third period, there was no large fluctuation except for the increases in \textsf{Email} and \textsf{WebPage}.

The study rates for the CS items showed a slow upward trend up to around 2019, with no distinction between the first and second periods.
From around 2020, in the middle of the second period, they began to rise more steeply, notably for \textsf{Programming}. An abrupt increase was seen in 2025 for almost all items.

The acquisition rates for the literacy items lacked a consistent trend: \textsf{WordProcessing} and \textsf{Email} kept declining, \textsf{Presentation} showed a slow increase,
and \textsf{Typing} decreased until around 2019 and then increased.
In all cases, however, the changes were slow. In the third period, slight upward trends were seen for \textsf{WebPage} and \textsf{MultiMedia}.

Among the acquisition rates for the CS items, \textsf{Ethics} was an exception. It was relatively high from the beginning, rose slightly around 2016, and remained constant thereafter. The other items were consistently low throughout the first and second periods but increased clearly and discontinuously in 2025.

\subsection{Comparison of DE and GY in 2025 with the Preceding and Following Years}
\begin{figure}[tb]
\centering
\noindent\begin{minipage}{\linewidth}
\centering
\includegraphics[height=.28\textheight]{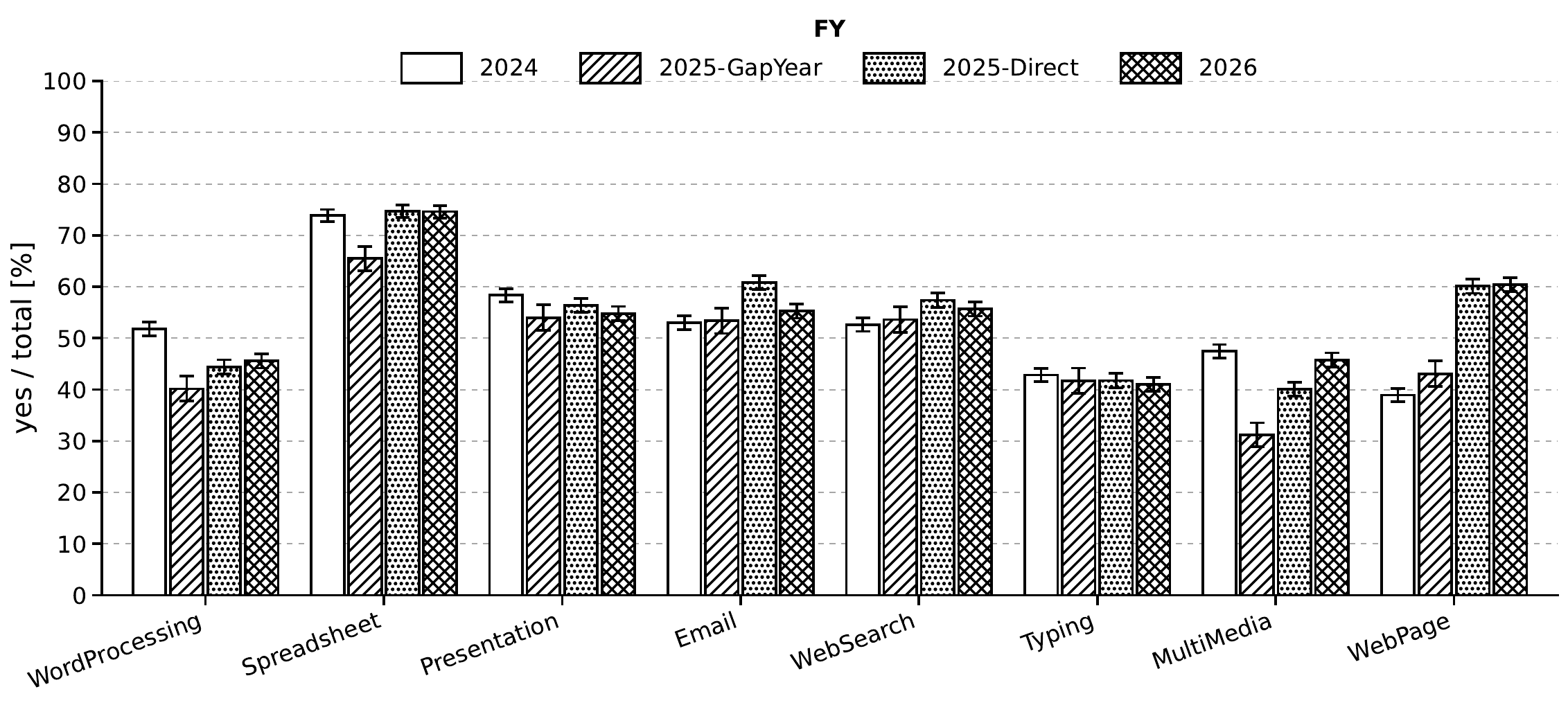}
\end{minipage}\\
\begin{minipage}{\linewidth}
\centering
\includegraphics[height=.28\textheight]{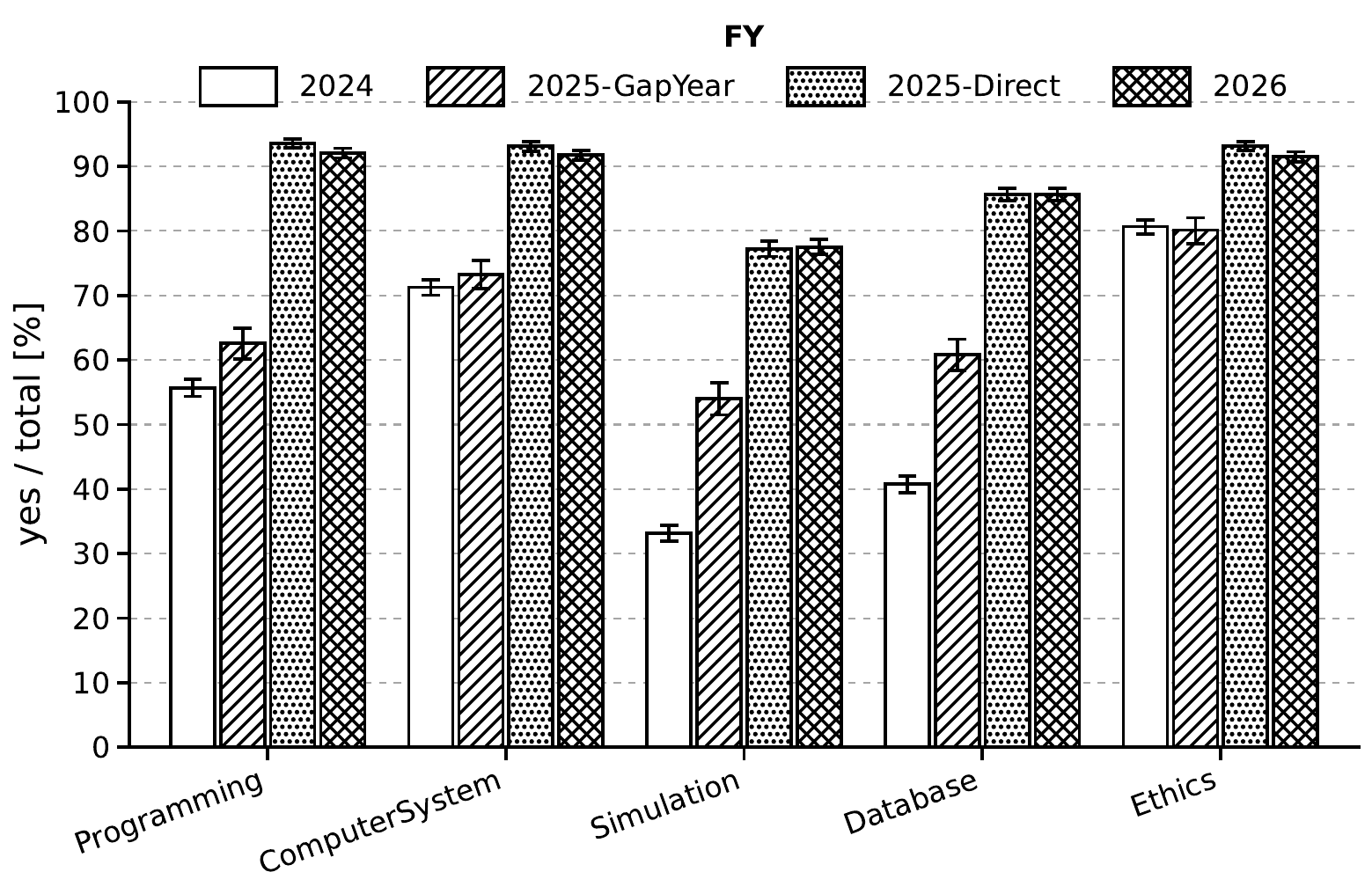}
\caption{DE-GY comparison of study rates: (upper) literacy items, (lower) CS items. Wilson score intervals with the finite population correction applied (target level 95\%) are shown.}\label{fig:comp1}
\end{minipage}
\end{figure}

\begin{figure}[tb]
\noindent\begin{minipage}{\linewidth}
\centering
\includegraphics[height=.28\textheight]{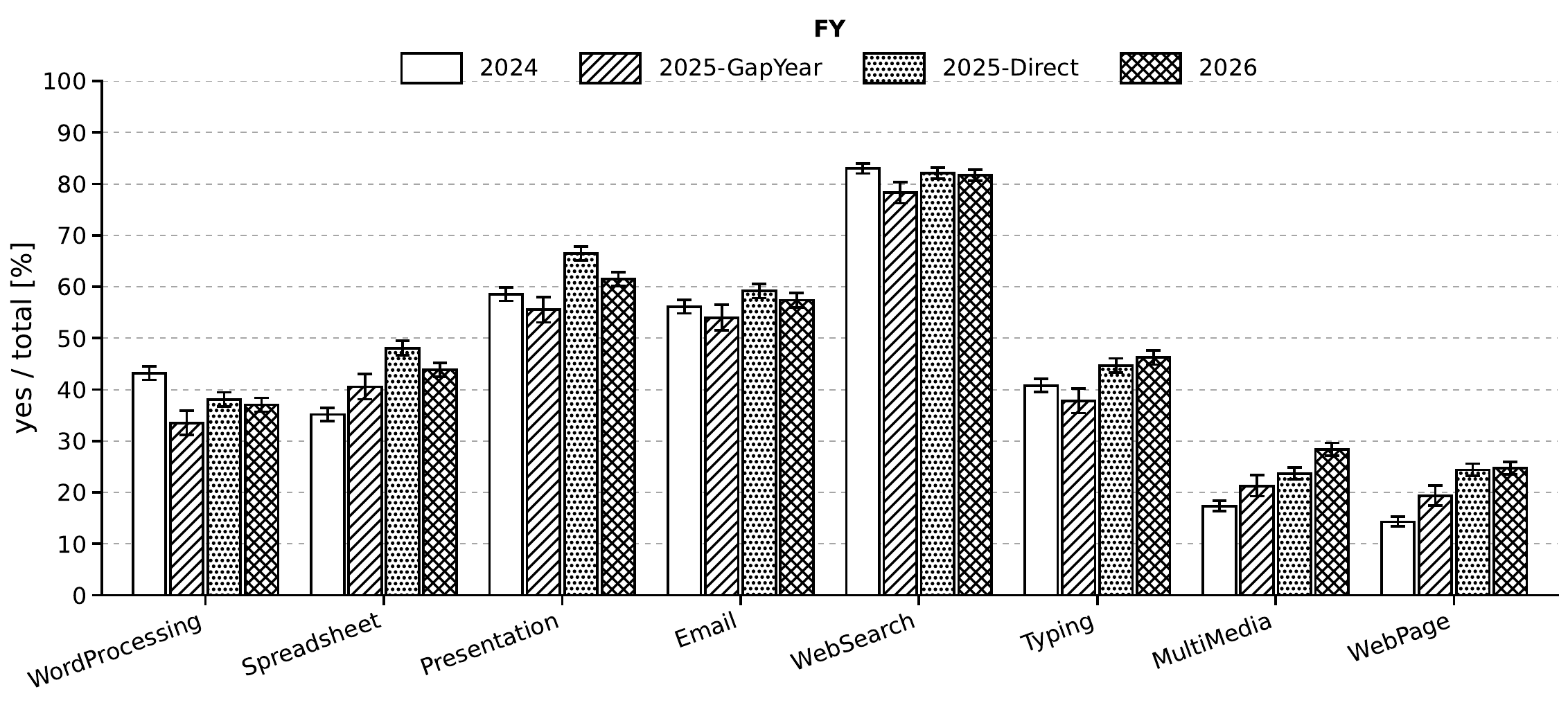}
\end{minipage}\\
\begin{minipage}{\linewidth}
\centering
\includegraphics[height=.28\textheight]{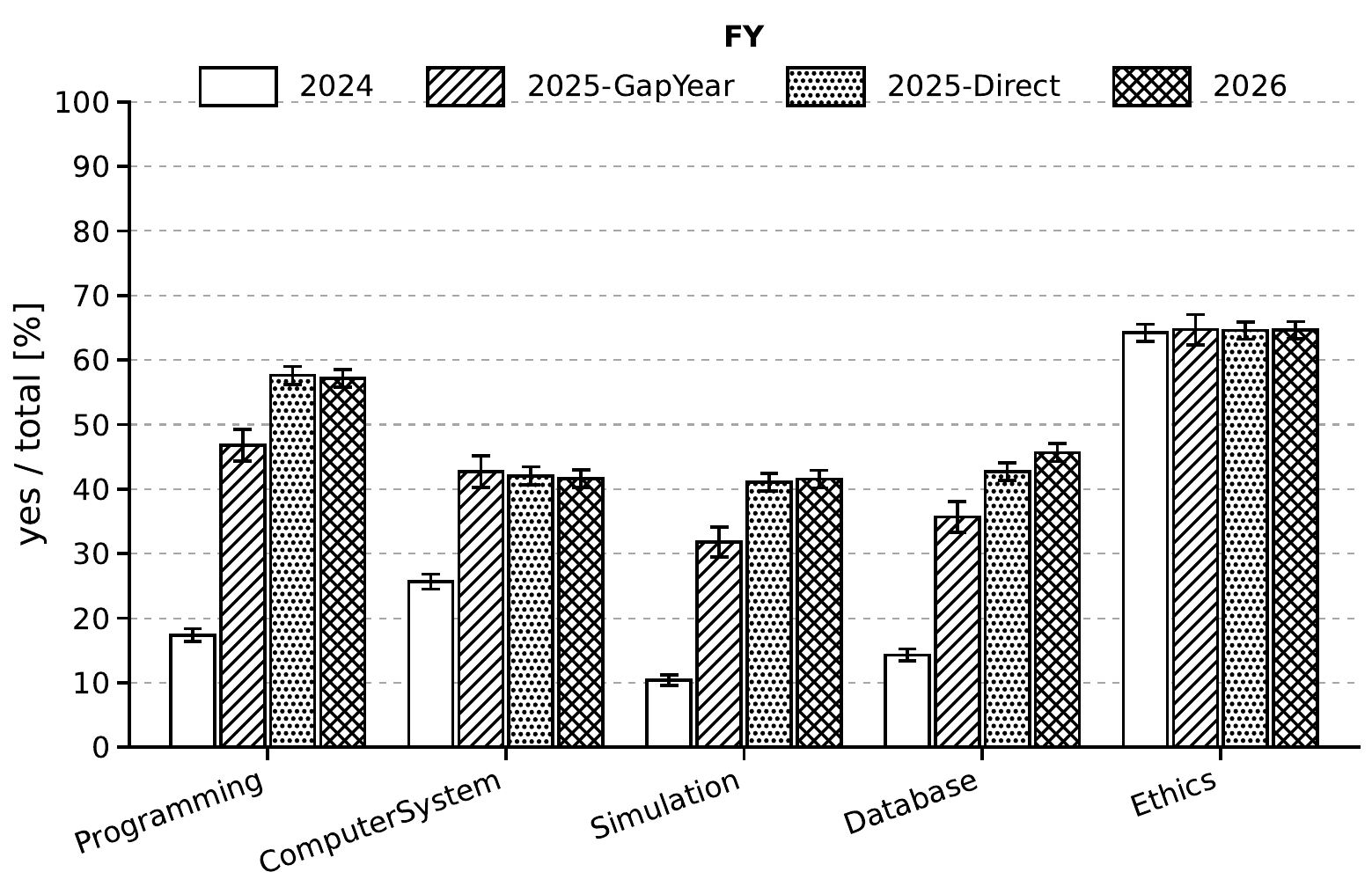}
\caption{DE-GY comparison of acquisition rates: (upper) literacy items, (lower) CS items. Wilson score intervals with the finite population correction applied (target level 95\%) are shown.}\label{fig:comp2}
\end{minipage}
\end{figure}
To analyze the impact of the CT in more detail,
we divided the students entering in 2025 into ``DE'' and ``GY'' and compared them with the students entering in 2024\footnote{Since GY entering in 2025 includes those who spent more than one year, it is natural to compare them with all students entering in 2024 rather than with DE alone.} and 2026.
Figures \ref{fig:comp1} and \ref{fig:comp2} show the results.

Compared with those of the students entering in 2024, the study rates of GY tended to be lower for some literacy items and higher for the CS items overall.
For the literacy items, the acquisition rates of GY were at a similar level to those of the students entering in 2024, except for \textsf{WordProcessing}.
For the CS items, the acquisition rates of GY were much higher than those of the students entering in 2024 and reached values comparable to those of DE, except for \textsf{Ethics}.

For DE, both the study rates and the acquisition rates were at the same level as those of the students entering in 2026.

\section{Discussion}
\subsection{RQ1: How have Japanese students' perceptions of the informatics field changed during the first period and the second period?}
First, we look at the trend in study rates from the first to the second period.

The most notable change is the one in the literacy items from an upward trend in the first period to a downward trend in the second period.
Nevertheless, the change remains continuous and slow. There is no visible gap between the first period and the second period. The clearest changes can be seen in \textsf{WordProcessing} and \textsf{Presentation}, but their study rates began to decline around 2019, not from 2016, the year when the second-period students began to enter.
The CS study rates again show no gap between the first and the second periods. In sum, the study rates in these periods follow a consistent trend. The curriculum reform from the first period to the second period does not appear to have had a major impact.

Then, what brought about the change in study rates?
If it was not the effect of the curriculum, it is reasonable to consider that the change occurred gradually, driven mainly by teachers. Since the importance of CS education had been recognized, the change makes sense if we regard it as a gradual improvement of instruction and materials by teachers. The acceleration in the increase of the CS study rates from around 2020 is worth noting. This is probably because teachers shifted their class content toward the CS fields in anticipation of the introduction of the CT. Note that the study rates of some literacy items (\textsf{WordProcessing} and \textsf{Presentation}, in particular) declined during this period.

Next, we look at the change in acquisition rates from the first period to the second period.

Somewhat surprisingly, acquisition rates are largely inconsistent with study rates.
As for literacy items, while study rates follow a fairly consistent trend, acquisition rates vary by item:
\textsf{Email} and \textsf{WordProcessing} declined,
\textsf{Presentation} improved, and \textsf{Typing} declined and then improved.
As for CS items, while the study rates rose gradually, the acquisition rates hardly improved.

What, then, affects acquisition rates? A possible interpretation is that what matters is whether a topic is experienced and acquired in students' daily lives. For example, the spread of images and videos has distanced students from the formal text documents of \textsf{WordProcessing}, and messaging applications have distanced them from \textsf{Email}.
In contrast, students have probably studied and acquired \textsf{Presentation} as part of ICT literacy since they were primary school students.
The influence of Covid-19 seems to have affected \textsf{Typing}, considering when the trend turned upward.
The online classes brought about by Covid required extensive use of chat and computer-based testing and made typing more familiar to students.

\paragraph{Answer to RQ1:}
From the first period to the second period, the content studied shifted slowly toward CS. Acquisition varied by item, but the change was not large overall. No direct impact of the curriculum reform from the first period to the second period can be observed. The change in the content studied appears to be influenced more by teachers' efforts at instructional improvement. The change in the content acquired is influenced more by how familiar the topics are to students.

\subsection{RQ2: Did the curriculum reform in 2022 and the introduction of the CT in 2025 produce a discontinuous change in students' perceptions of the informatics field?}\label{subsect:RQ2}
Next, we analyze the change to the third period, comparing it with the change to the second period.

For the literacy items, neither the study rates nor the acquisition rates changed much in the third period.
There are a few exceptions: \textsf{Email} and \textsf{WebPage} increased in study rates, and \textsf{WebPage} and \textsf{MultiMedia} in acquisition rates. We discuss these exceptions later.

The study rates and acquisition rates for the CS items increased abruptly in the third period, except for \textsf{Ethics}. This change is too large, compared with the change to the second period, to be interpreted as the impact of the curriculum reform alone. It is therefore natural to interpret it as a washback effect caused by the introduction of the CT. The fact that this change does not appear in the literacy items also supports this interpretation. 

However, this result is difficult to take at face value. Is it really true that, for \textsf{Programming}, which less than 20\% of the students had ``acquired'' in 2024, about 55\% of the students ``acquired'' it in 2025? Compared with the other items, this would mean that \textsf{Programming} is easier than \textsf{Typing} and \textsf{WordProcessing} and about as easy as \textsf{Presentation} and \textsf{Email}. These ``facts'' are clearly counter-intuitive.

We closely examine this change through the lens of the washback effect.
The change in study rates is consistent with previous findings that teachers rapidly adapt their class content in response to a change in high-stakes tests \parencite{Alderson93,Cheng98,Cheng04}.
On the other hand, regarding the change in acquisition rates, the previous findings support the intuition that it is unreasonable to interpret the change as an improvement in acquisition.
Previous studies have found that the washback effect on learners' products is not large \parencite{Alderson93,Cheng04,Pan14}.

Then, which aspects of learners are known to be directly susceptible to washback?
Previous studies point to learners' subjective perceptions of the test and of the field \parencite{Li12, Xie13, Pan14, Xie25}: motivation and attitude toward the test, the perceived difficulty of the test, and self-efficacy.
Given this, it is more reasonable to interpret the change as \emph{a change in the criterion for judging the acquisition of a topic}.
For example, in the case of \textsf{Programming}, the students before 2024 may have thought that they could hardly say ``acquired'' unless they could write programs at a professional level, whereas from 2025 they may have felt that they had ``acquired'' it once they could solve the CT questions.

This interpretation also explains the change in the acquisition rate of \textsf{Ethics}.
Like the other items in the CS domain, the study rate of \textsf{Ethics} rose further in the third period; however, it is the only CS item whose acquisition rate did not rise.
This is rather strange, given that \textsf{Ethics} is a typical topic examined in the CT and that its acquisition rate was about 64\%, relatively high but not close to the ceiling.
We can account for this anomaly if we interpret it as a case in which the criterion for judging acquisition remained unchanged.
The elementary content of information ethics is close to common sense, such as ``privacy is important'' and ``copyright must be respected''. Students may have felt that understanding that much was sufficient. In preparing for the entrance examination, however, students are required to understand more advanced content, such as what is protected by the Copyright Act and the Act on the Protection of Personal Information, and how cryptographic techniques are used in information security. As a result, many students probably felt that the topic was more difficult than they had thought.

Finally, we discuss \textsf{Email}, \textsf{WebPage}, and \textsf{MultiMedia}. Although their increases are not as large as those of the CS items, they are too large to be regarded as a result of the curriculum reform. They may reflect a washback effect.
``Informatics I'' includes, besides the CS domain, a learning topic called information design.
Because it is rather difficult to create exercises for information design, many textbooks and workbooks took up topics for which materials are relatively easy to prepare, such as email etiquette, building websites with HTML, and compression schemes for multimedia data. We suppose that this is reflected in the increases.

\paragraph{Answer to RQ2:}
The change to the third period is far greater than the change to the second period, suggesting a substantial washback effect of the CT. Moreover, its impact is too large to be interpreted as an improvement in students' proficiency. Instead, it seems to stem largely from a change in students' criterion for judging acquisition.

\subsection{RQ3: To what extent do perceptions of the informatics field differ between students who followed the same curriculum but were subject to different entrance examination systems, and between those entering in the same year but following different curricula?}\label{subsect:RQ3}
First, we compare GY and the students entering in 2024. They followed the same curriculum and differ only in whether they took the CT.

The change in study rates must be noted\footnote{For the five items with the largest changes, \textsf{WordProcessing}, \textsf{SpreadSheet}, \textsf{MultiMedia}, \textsf{Simulation}, and \textsf{Database}, the difference is significant at the 5\% level in a chi-square test with the Bonferroni correction (13 comparisons).}. Since these students studied under the same curriculum, the actual study rates ought to be the same. Moreover, the directions of the changes are consistent with the category of the item. The literacy items declined and the CS items increased. These changes are therefore likely to reflect a washback effect that affected students' perceptions.

GY students studied mainly CS content while preparing for the entrance examination. As a result, their impression of having ``studied'' CS might have been reinforced---or they might ``discover'' that high school Informatics covers such content. Similarly, their impression of ``Literacy'' might have faded---or they might come to misunderstand that such content is not part of Informatics in high school.

Next, we look at acquisition rates. For the literacy items, the difference between GY and the students entering in 2024 is small compared with that in study rates, but
\textsf{WordProcessing} again declined\footnote{The difference is significant at the 5\% level in a chi-square test with the Bonferroni correction (13 comparisons).}. For the CS items,
all except \textsf{Ethics} increased abruptly. Their values are even closer to those of DE than to those of the students entering in 2024. It is natural to interpret this difference, too, as a washback effect.

Note that the study rates and acquisition rates of GY for some items, all in literacy, are clearly lower than those of the students entering in 2024.
Since GY should have studied Informatics again before the entrance examination in order to prepare for the CT, it would be more natural for these values to rise.
A reasonable interpretation is negative washback on the topics unlikely to be examined in the CT. This is consistent with the study by \textcite{Scanlon21} and with the findings of previous washback studies \parencite{Cheng04,Qi07,Pan14}.

Second, we compare GY and DE. For study rates, DE tend to be higher overall, notably for the CS items\footnote{For the nine items other than \textsf{WordProcessing}, \textsf{Presentation}, \textsf{WebSearch}, and \textsf{Typing}, the difference is significant at the 5\% level in a chi-square test with the Bonferroni correction (13 comparisons).}.
Acquisition rates are also higher for DE overall\footnote{For the four items \textsf{SpreadSheet}, \textsf{Presentation}, \textsf{Programming}, and \textsf{Simulation}, the difference is significant at the 5\% level in a chi-square test with the Bonferroni correction (13 comparisons).}, but the difference is smaller than that in study rates.
Among these differences, those in study rates for the CS items are natural and consistent with the curriculum reform. The differences in acquisition rates for the CS items are also understandable, given that they are not so large, because DE studied the content more thoroughly according to the curriculum.

It may seem strange that GY are lower than DE for some literacy items in both study rates and acquisition rates.
Since DE should have put more effort into the CS domain in their high schools, one would expect their literacy to be weaker.
Again, this suggests negative washback on the topics unlikely to be examined in the CT. Because GY spend a full year devoted to preparing for the entrance examination, the washback effect seems to have appeared more noticeably, both positively and negatively, in GY than in DE.

\paragraph{Answer to RQ3:} Study rates and acquisition rates are influenced mainly by the curriculum and the CT, respectively. However, the CT also affects study rates, which strongly suggests a change in students' perceptions.
The washback effect is positive for the CS domain but somewhat negative for literacy.

\subsection{Implications for Informatics Education}\label{subsect:imply}
The results of the analysis in this paper may have the following implications for informatics education.

First, curriculum reform alone did not have a direct or significant impact on students' learning.
This does not mean that curriculum reform is meaningless; rather, its effects appear only slowly.
Meanwhile, the degree of students' acquisition is strongly influenced by what is familiar to them.
Since the ICT environment and devices surrounding students shift rapidly,
measuring the effect of curriculum reform is likely to be difficult.

Second, the impact of a high-stakes test on students is drastic. The change it produced in a single year exceeded the accumulated change over nearly 20 years for the CS domain. However, what changed and to what extent is not obvious.
Indeed, one of the findings of this study is the difficulty of measuring this.

This study demonstrated that measurement by a test, a high-stakes test in particular, is inherently bounded by the areas it covers. For the areas a test covers, it may provide accurate measurement. However, it brings about a concentration on the examined content and thereby probably leads to a negative washback effect in the other areas.
This issue is important in informatics education, because the field includes areas, such as literacy, for which paper tests are not suitable. For small- to medium-scale measurement, adding performance tasks, as in AP CSP, is a natural aid. For a large-scale high-stakes test such as the CT in Japan, a solution is not obvious.

Students' self-reports of their acquisition have also been widely used.
However, this study revealed the low reliability of such self-reports as a measurement of proficiency.
In particular, across an institutional modification (in either the curriculum or the examinations), the criterion on the students' side may change, which undermines comparison over time.
In this study, by separating DE and GY and partially controlling for influences other than the CT, we were able to detect a change in the criterion for reporting acquisition. Even so, we do not know how much the criterion changed, or how far it is consistent with those of teachers and of test and curriculum designers.

Nevertheless, in informatics education, there is a considerable reason to use students' self-reports of their acquisition as a measurement. Given the rapid change in the situation surrounding ICT, there is little point in comparing the proficiency of current students with that of students 20 years ago. For example, now that large language models have become ubiquitous, measuring \textsf{WebSearch} skills is becoming outdated.

The discussions above can be summarized as the following two recommendations. First, when a high-stakes test is administered,
we should prepare continuous measurement for the areas that are not examined.
Second, if the purpose is to measure proficiency, self-reports of acquisition should not be interpreted literally, especially across an institutional change. They should be combined with other measurement methods to detect and calibrate changes in the criterion. These recommendations are particularly important for Informatics, which includes both the CS domain and the literacy domain.

Finally, in this study we analyzed DE and GY separately in order to isolate the impacts of the CT from the curriculum reform. This approach can be used to separate the effects of a curriculum and an entrance examination
if there is a sufficient number of students whose year of high school entry differs from that of the majority.

\subsection{Limitation}\label{subsection:limitation}
This study has several inherent limitations.

First, the present data capture only students' perceptions, not their proficiency.
As discussed in Section \ref{subsect:imply}, this is hard to avoid in a longitudinal study, but
it would be desirable for future studies to include proficiency in the analysis.

As discussed in Section \ref{subsect:bias}, the present data have non-response bias. Even worse, the years with relatively large bias are 2016 (when the second-period students began to enter), 2025, and 2026, which are precisely the years this study focuses on.
Nevertheless, we expect the impact of the bias on the present results to be marginal.
The survey results are largely consistent over the years.
In particular, 2016 is consistent with the results of the preceding and following years. The change in 2025 is so large that a small amount of non-response bias could not alter the overall picture. That said, the ``improvement'' in study and acquisition rates may be somewhat overestimated because of the bias in the respondents toward STEAM in 2025 and 2026.

As noted in Section~\ref{subsect:survey}, the wording of some items was changed in 2025. Up to 2024, we presented only the item names in Table \ref{table:items1} and asked whether students had studied and/or acquired them. In 2025, for the five CS items, \textsf{Ethics}, \textsf{Programming}, \textsf{ComputerSystem}, \textsf{Simulation}, and \textsf{Database}, supplementary keywords representing the content of each item were added, and \textsf{MultiMedia} was modified from ``image processing and multimedia'' to ``image processing and multimedia \emph{software}''. These modifications may have changed students' impressions of the items. For example, for \textsf{Programming}, keywords such as ``conditional branching'' and ``iteration'' might have given students the impression that their understanding was sufficient. We also pointed out that the acquisition rate of \textsf{Ethics} did not rise; keywords such as ``copyright and intellectual property rights'' might have given an intimidating impression.

In this regard, \textcite{Yama26} compared study rates for GY entering in 2025 and DE entering in 2024, observed that the ratios of study rates for the items whose wording was modified varied (0.66--1.56) whereas the ratios of study rates for others were close to 1 (0.90--1.11), and conjectured that the differences in study rates were due to these modifications.

In this study, we do not attribute the differences between GY entering in 2025 and students entering in 2024 to the wording modifications. The reason is that, as discussed in Section \ref{subsect:RQ3}, \textsf{WordProcessing}, \textsf{SpreadSheet}, \textsf{MultiMedia}, \textsf{Simulation}, and \textsf{Database} showed statistically significant differences. The change in wording does not account for the differences for \textsf{WordProcessing} and \textsf{SpreadSheet}, which were not modified, nor does it explain the large gap for \textsf{MultiMedia}, whose modification was minimal. Nevertheless, the effect of the CT cannot be completely isolated from that of the wording modifications. 

The University of Tokyo is generally regarded as the most competitive university in Japan in terms of entrance examinations. This brings about a selection bias toward students who studied Informatics seriously in high school.
At the same time, many of the students entering the University of Tokyo come from high-performing schools, and this can also lead to bias.
Previous studies \parencite{Zhang21,Dong22} point out that these biases produce systematic differences in the washback effect. In high-performing schools, teachers actively carry out instructional improvement based on a long-term perspective, and students rate the importance and validity of the test highly and eagerly prepare for the test.
This is consistent with the results of this study: the introduction of the CT promoted instructional improvement, and students adopted the ability to solve examination questions as their criterion for acquisition.
In other words, because the respondents are students entering the University of Tokyo, the washback effect may have been observed more clearly than it would be elsewhere.

We attributed the differences in study and acquisition rates to the entrance examination because an effect of the curriculum reform is unlikely. However, we cannot conclude that the differences between years are due only to the Courses of Study and the entrance examination. For example, GY may be affected by the fact that one more year has passed since they studied Informatics than in the case of DE. The differences may also reflect social changes such as the progress of AI technologies. In this paper, we assume that the other influences are relatively small, because the year-to-year variation of this survey is generally not large and because, for students who enter the University of Tokyo, the entrance examination matters more than almost anything else. This is a limitation of this paper.

The author teaches informatics to first-year university students. This circumstance of the author may have been a source of bias in the analysis. The author must design the classes on the assumption that the institutional reform did not improve students' proficiency dramatically. Accordingly, the author may tend to think that the effect of the entrance examination reform is limited. The author's impression has not changed much even after teaching the course to the students who entered in 2025 and 2026. Even so, this is also a limitation of this paper.

\section{Conclusion}
In this paper, we examined the impact of curriculum reform and the introduction of a large-scale entrance examination in the informatics field. We analyzed a questionnaire administered longitudinally for two decades at the University of Tokyo through the lens of washback effects.  As a result, we found the following three points.
\begin{itemize}
\item Over these two decades, while the content studied shifted gradually toward the CS domain, the content students acquired did not change much. In particular, the curriculum reforms had a limited impact on students' study and acquisition.
\item The introduction of the CT had a large impact. In particular, its effect on the study rates and acquisition rates for the CS items was significant.
\item The change in students' ``acquisition rates'' most likely derived from a change in the criterion for judging ``acquired'' topics.
\end{itemize}

To the author's knowledge, this study is the first authentic application of the framework of the washback effect to informatics education. It will serve as a starting point for further deepening the discussion of student assessments in informatics education.

\printbibliography

\end{document}